\documentclass[10pt,leqno]{amsart}
\usepackage{titlesec}
\usepackage{amsfonts}
\usepackage{amssymb}
\usepackage{amsmath}
\usepackage{amsthm}
\usepackage{graphicx}
\usepackage{fancyhdr}

\usepackage{indentfirst,csquotes}

\usepackage{amssymb,amsthm,amsmath}
\usepackage{xcolor,paralist,hyperref,titlesec,fancyhdr,etoolbox}

\hypersetup{ colorlinks=true, linkcolor=black, filecolor=black, urlcolor=black }

\usepackage{lipsum}

\titleformat{\section}[block]{\normalfont\Large\bfseries\centering}{\thesection.}{1em}{}

\begin{document}

\title{Computing Fundamental Constants in the FLRW Universe using the Hawking Radiation of the Cosmological Horizon} 
\author[Initial Surname]{A. Meza$^1$ and P. Padilla$^2$}
\date{\today}

\footnotetext[1]{Escuela de Ingeniería y Ciencias (EIC), Instituto Tecnológico y de Estudios Superiores de Monterrey (ITESM), Monterrey, Nuevo León, México. \\ armando\_meza\_g@tec.mx}
\footnotetext[2]{Instituto de Investigaciones en Matemáticas Aplicadas y en Sistemas (IIMAS), Universidad Nacional Autónoma de México (UNAM), Ciudad de México, México. \\ pablo@mym.iimas.unam.mx}

\let\thefootnote\relax

\footnote{MSC2020: Primary 00A05, Secondary 00A66.}

\maketitle

\begin{abstract}
In this work, we compute the universe temperature for the cosmological horizon for the FLRW metric. For this purpose, we consider a scalar field on the cosmological horizon. This scalar field satisfies the Klein-Gordon equation in a curved space-time. Recently, some authors like Barrow, Bekenstein and others have proposed that the fundamental constants might vary with time. Along these lines of thought, we derive the electromagnetic momentum and electromagnetic field for the FLRW universe. This enables us to obtain a temporal dependence of the Hubble parameter which, in turn, induces a time dependence of the fundamental constants. In order to validate this model, the theoretical predictions are then compared with observational data as a function of the redshift, $z$, for the universe expansion. As a consequence, this time dependence on the fundamental constants makes it possible to predict a change in time of the informational content for the entropy of the universe surface area, this is expressed as the bit number by using the holographic principle. 
\end{abstract} 


\section{Introduction}

In recent years the possibility has been considered by various authors (\cite{Barrow}, \cite{Barrow_1}, \cite{Bekenstein_1}, \cite{Bekenstein_2}, \cite{Damour_1}, \cite{Damour_2}, \cite{Khoury_1}, \cite{Khoury_2} and \cite{Mota}) that the fundamental constants should change in time with the expansion of the universe. This problem about the variation of fundamental constants has allowed multiple studies based on the hypothesis of large Dirac numbers \cite{Dirac_1}, \cite{Dirac_2}, \cite{Dirac_3}. Even the theories of quantum gravity and the attempts to unify the four fundamental forces take into consideration the study of the temporal variation of the fundamental constants \cite{Barrow}, \cite{Barrow_1}, \cite{Bekenstein_1}, \cite{Bekenstein_2}, \cite{Damour_1}, \cite{Damour_2}, \cite{Khoury_1}, \cite{Khoury_2} and \cite{Mota}.

Different physical aspects that have been taken into account when analyzing the variation of the fine structure constant are covariance, gauge invariance, and causality (invariance in the face of temporal reversals). One of the few ways to contrast unification theories is the study of the variation of the fundamental constants. This is because the standard model of particle physics and general relativity involve certain parameters that have the particularity of remaining invariant in space-time and in a reference system, these parameters are the fundamental constants \cite{Barrow}, \cite{Barrow_1}, \cite{Bekenstein_1}, \cite{Bekenstein_2}, \cite{Damour_1}, \cite{Damour_2}, \cite{Khoury_1}, \cite{Khoury_2} and \cite{Mota}.

In 1937 Dirac formulated the hypothesis that the fundamental constants are a function of the age of the universe, that is, these constants can present spatial and temporal variation with respect to the expansion of the universe \cite{Dirac_1}, \cite{Dirac_2}, \cite{Dirac_3}. The standard model and general relativity have different predictions for the variation of these parameters \cite{Barrow}, \cite{Barrow_1}, \cite{Bekenstein_1}, \cite{Bekenstein_2}, \cite{Damour_1}, \cite{Damour_2}, \cite{Khoury_1}, \cite{Khoury_2} and \cite{Mota}. These parameters are not determined by theories but are the result of experimental measurements \cite{Barrow}, \cite{Barrow_1}, \cite{Bekenstein_1}, \cite{Bekenstein_2}, \cite{Damour_1}, \cite{Damour_2}, \cite{Khoury_1}, \cite{Khoury_2} and \cite{Mota}.

Dirac's observation to formulate his hypothesis of large numbers was that the order of magnitude of the quotient between the electromagnetic and gravitational forces between a proton and an electron coincided with the ratio between the age of the universe and the time it takes for light to pass through a hydrogen atom $(~10^{40})$ \cite{Dirac_1}, \cite{Dirac_2}, \cite{Dirac_3}.

In this sense, quantum gravity theories such as superstring theories and M theory predict the variation of fundamental constants in space-time; this can be studied experimentally only at low energies \cite{Sherrill} and \cite{Nieto}. 

Some of these theories are Bekenstein's theories for the fine structure constant $\alpha$ (\cite{Bekenstein_1}, \cite{Bekenstein_2}). Also included are theories where the speed of light $c$ is variable in time and space \cite{Albrecht}, \cite{Barrow_3}, \cite{Moffat_1} and \cite{Moffat_2}.

In these theories, the constant whose variation is to be studied is replaced by a scalar field responsible for this variation. The prediction of the variation of the constants is deduced from the dynamics of the field. In the case of Bekenstein's theory, the variation of the fine structure constant $\alpha$ is studied based on the variation of an action that is a function of the Planck mass $m_{pl}$ and a scalar field $\phi$ \cite{Bekenstein_1}, \cite{Bekenstein_2}.

On the other hand, it has recently been found in \cite{Ya-Peng} that for the FLRW metric, it is possible to consider the cosmological horizon. In this cosmological horizon, the temperature belonging to Hawking radiation can be measured. For this, it was considered that near the cosmological horizon, there is the presence of a scalar field described by the Klein-Gordon equation in curved space-time for the FLRW metric. The solution of this Klein-Gordon equation is also the solution of the wave equation. From this wave equation, we can understand that near the horizon there may be incoming or outgoing waves that represent the creation and destruction of particles and antiparticles. By having this, the strength of the incoming and outgoing waves for the horizon was considered, with which the states of the particles and antiparticles were obtained, this allowed, using a scalar product, to find a thermal spectrum similar to that of a black body and with it find the temperature belonging to the Hawking radiation for the cosmological horizon; this temperature was found based on the Hubble parameter \cite{Ya-Peng}.

It is also possible thanks to the study of the holographic principle and the $AdS/CFT$ correspondence that has been formulated by several authors: \cite{Bekenstein_3}, \cite{Bekenstein_4}, \cite{Bekenstein_5}, \cite{Bekenstein_6}, \cite{Hawking_1}, \cite{Hawking_2}, \cite{Hawking_3}, \cite{Hawking_4}, \cite{Hooft_1}, \cite{Hooft_2}, \cite{Susskind_1}, \cite{Susskind_2}, \cite{Susskind_3}, \cite{Susskind_4}, \cite{Maldacena}, \cite{Wheeler}, \cite{Unruh} and \cite{Gibbons} among others. The temperature of black holes given by Hawking radiation \cite{Hawking_2} can be associated with its entropy which is a function of the surface area of the above mentioned black hole. In the holographic principle, it is also stated that for every four Planck units there is at least one degree of freedom (or a constant Boltzmann unit $k$ of maximum entropy). This is known as the Bekenstein boundary $S < A/4$ \cite{Bekenstein_4}, \cite{Bekenstein_5}, \cite{Bekenstein_6}. In a broader and more speculative sense, the theory suggests that the entire universe can be viewed as a two-dimensional information structure "painted" on the cosmological horizon, such that the three dimensions that are observed would only be an effective description at macroscopic scales and low energies; so the universe would actually be a hologram \cite{Susskind_1}, \cite{Susskind_2}. In that sense, Wheeler suggested that the number of bits for the universe is $N=8 \times 10^{88}$ \cite{Wheeler}, this quantity will be considered in our calculations.

Now, the QED of the standard model is a gauge theory that allows us to describe the electromagnetic interaction \cite{Feynman_1}, \cite{Feynman_2}, \cite{Feynman_3}, \cite{Dirac_4}, \cite{Fermi}, \cite{Oppenheimer}, \cite{Schwinger_1}, \cite{Schwinger_2}, \cite{Schwinger_3} and \cite{Dyson}. Thanks to this theory, it is possible to compute the Euler-Lagrange equations from the Klein-Gordon action in curved space-time coupled to the electromagnetic field \cite{Beltran}. It has previously been possible to find the Ward-Takahashi equation for the 4-momentum of the electromagnetic field for the universe which must be solved numerically \cite{Ward}, \cite{Takahashi} and \cite{Peskin}.

In order to consider the contribution of the electromagnetic field to the Hubble parameter, in this paper it will be seen that it is possible to take into account this parameter as a function of the 4-momentum for the electromagnetic field. This depends on the electromagnetic potential for the radius of the universe (on the cosmological horizon). By obtaining this factor, it is thus possible to use the expression for the temperature in the cosmological horizon given by \cite{Ya-Peng} as a function of the expansion of the electromagnetic field given by the expansion of the universe.

Thanks to the above and considering the hypothesis of large Dirac numbers \cite{Dirac_1}, \cite{Dirac_2}, \cite{Dirac_3}, an expression will be obtained for the Hubble parameter as a function of the temperature of the universe. Various fundamental constants that arise from the electromagnetic interaction, such as the fine structure constant $\alpha$, the electron charge $e$, the electron mass $m_e$, the redshift $z$, the speed of light $c$ and the permeability of the vacuum $\epsilon_0$ will be considered. From these values it will be possible to address the change in the fundamental constants depending on the expansion of the universe \cite{Dirac_1}, \cite{Dirac_2}, \cite{Dirac_3}. The latter can be analyzed as a case study using astronomical data thanks to the concept of apparent magnitude.

In section 2, a brief review will be given about how Ya-Peng Hu \cite{Ya-Peng} calculated the temperature from the Hawking radiation for the cosmological horizon for the Klein-Gordon equation in curved space-time. This will be compared with the Bekenstein-Hawking entropy and the holographic principle to give an expression for the number of bits in the universe \cite{Bekenstein_3}, \cite{Bekenstein_4}, \cite{Bekenstein_5}, \cite{Bekenstein_6}, \cite{Hawking_1}, \cite{Hawking_2}, \cite{Hawking_3}, \cite{Hawking_4}, \cite{Wheeler}, \cite{Unruh} and \cite{Gibbons}.

In section 3, by considering the theory of QED \cite{Feynman_1}, \cite{Feynman_2}, \cite{Feynman_3}, \cite{Dirac_4}, \cite{Fermi}, \cite{Oppenheimer}, \cite{Schwinger_1}, \cite{Schwinger_2}, \cite{Schwinger_3} and \cite{Dyson}, the 4-momentum of the electromagnetic field of the universe, the hypothesis of large Dirac numbers \cite{Dirac_1}, \cite{Dirac_2}, \cite{Dirac_3}, the expansion of the universe, the temperature of the cosmological horizon found by Ya-Peng Hu \cite{Ya-Peng}, and the holographic principle \cite{Bekenstein_3}, \cite{Bekenstein_4}, \cite{Bekenstein_5}, \cite{Bekenstein_6}, \cite{Hawking_1}, \cite{Hawking_2}, \cite{Hawking_3} and \cite{Hawking_4}, it will be possible to find an expression for the Hubble parameter \cite{Hubble} as a function of the temperature and radius of the universe, the number of bits \cite{Wheeler} contained in it and some fundamental constants, taking special interest in the fine structure constant $\alpha$ \cite{Dirac_1}, \cite{Dirac_2}, \cite{Dirac_3}. This is done with the objective of understanding that the expansion of the universe contributes to the variation of the fundamental constants and, in turn, to the variation of the temperature and entropy of the universe. This is reflected then in the fact that there may be a relationship between the number of bits $N$ and the variation of the fundamental constants, two important objects of study in quantum gravity theories that attempt to unify all fundamental forces.

In section 4, a comparison is made between the change in the apparent magnitude $\mu (z)$ with respect to the redshift $z$, making use of the Hubble parameter found in section 3 with observational data. This will be contrasted with Barrientos et al. work \cite{Barrientos}. On the other hand, it is possible a comparison between the Hubble parameter $H$ and the fine structure constant $\alpha$.  

In section 5 we discuss our results, while section 6 addresses the conclusions and final comments.

\section{The Cosmological Horizon and its Hawking Radiation in an FLRW Universe}

The observable universe, the horizon of the universe, and the cosmological horizon constitute the visible part of the total universe, the global geometry of the space-time of the universe appears to be flat. It has a radius of $4.40 \times 10^{26}$ m ($46.5$ billion light years), a volume of $1.08 \times 10^{79}$ $m^{3}$ and a mass of $9.27 \times 10^{52}$ kg, so the equivalent mass-energy density is $8.58 \times 10^{-27}$ $kg/m^{3}$. The word observable in this sense does not refer to the ability of modern technology to detect light or other information from an object, nor to whether there is anything to detect. It refers to the physical limit created by the speed of light itself. No signal can travel faster than light, so there is a maximum distance (called the particle horizon) beyond which nothing can be detected, since the signals could not have reached us yet \cite{Ya-Peng}.

A distinction is made between the visible universe, which includes only signals emitted since recombination (when hydrogen atoms were formed from protons and electrons and photons were emitted), and the observable universe, which includes signals from the beginning of cosmological expansion (the Big Bang in traditional physical cosmology, the end of the inflationary epoch in modern cosmology) \cite{Ya-Peng}.

According to the calculations carried out, the current comoving distance to the particles from which the Microwave Background Radiation $(CMBR)$ was emitted, which represents the radius of the visible universe, is about $14$ billion parsecs (about $45.7$ billion light years). The comoving distance to the edge of the observable universe is about $14.3$ billion parsecs (about $46.6$ billion light years), approximately  $2$ times greater. Therefore, the radius of the observable universe is estimated to be about $46.5$ billion light-years \cite{Ya-Peng}.

Currently, the limit of the observable universe is the cosmic microwave radiation that is at a redshift distance of $z=1089$, which means that we can see the universe since it was only $380$ thousand years old. If the cosmic neutrino background or the stochastic gravitational wave background were detected, the limit of the observable universe would extend to redshift distances above $z>1010$ or to a few fractions of a second after the Big Bang \cite{Ya-Peng}.

Also, the particle horizon (also called the cosmological horizon, the light horizon, or the cosmic light horizon) is the maximum distance that particles could have traveled toward the observer during the age of the universe. It represents the boundary between the observable and unobservable regions of the universe, so its distance at the present time defines the size of the observable universe. The existence, properties, and importance of a cosmological horizon depend on the particular cosmological model \cite{Ya-Peng}.

On the other hand, quantum field theory in curved space-time is an extension of standard quantum field theory in which the possibility that the space-time through which the field propagates is not necessarily flat is contemplated (described by the Minkowski metric). A generic prediction of this theory is that particles can be generated due to time-dependent gravitational fields, or the presence of horizons \cite{Ya-Peng}.

One of the consequences of Heisenberg's uncertainty principle is quantum fluctuations in the vacuum. These consist of the creation, for very brief moments, of particle-antiparticle pairs from a vacuum. These particles are "virtual", but the intense gravity of the black hole transforms them into real ones. Such pairs quickly disintegrate, returning the energy borrowed for their formation. However, at the edge of a black hole's event horizon, the probability that one member of the pair forms from the inside and the other from the outside is not zero, so one of the components of the pair could escape from the black hole. If the particle manages to escape, energy will come from the black hole. That is, the black hole must lose energy to compensate for the creation of the two particles separated. This phenomenon has the consequences of the net emission of radiation by the black hole and the decrease in its mass. Furthermore, a black hole emits thermalized Hawking radiation, according to a distribution identical to that of the black body corresponding to a temperature $T_{H}$, which expressed in terms of Planck units, turns out to be $T_{H}=(\alpha_H)/(2 \pi)$ \cite{Ya-Peng}.

Ya-Peng Hu considered the FLRW metric as follows \cite{Ya-Peng}:

\begin{equation}
ds^{2}=-dt^{2}+a^{2}(t)\left( \frac{d\rho ^{2}}{1-k\rho ^{2}}+\rho
^{2}d\Omega _{2}^{2}\right), \label{FLRW} 
\end{equation}
where $t$ is the cosmic time, $\rho $ is the comoving radial coordinate, $a$
is the scale factor, $d\Omega _{2}^{2}$ denotes the line element of a $2$
-dimensional sphere with unit radius, $k=1$, $0$ and $-1$ represent a
closed, flat and open FLRW universe respectively \cite{Ya-Peng}.

For convenience, Ya-Peng Hu defined $r=a\rho $. Therefore, the metric (\ref{FLRW}) can be rewritten as \cite{Ya-Peng}

\begin{equation}
ds^{2}=-\frac{1-r^{2}/r_{A}^{2}}{1-kr^{2}/a^{2}}dt^{2}-\frac{2Hr}{%
1-kr^{2}/a^{2}}dtdr+\frac{1}{1-kr^{2}/a^{2}}dr^{2}+r^{2}d\Omega _{2}^{2},
\end{equation} \label{FLRWA}
where $r_{A}=1/\sqrt{H^{2}+k/a^{2}}$ is the apparent horizon in an FLRW universe \cite{Ya-Peng}.

The metric of the de Sitter spacetime used by Ya-Peng Hu is \cite{Ya-Peng}

\begin{equation}
ds^{2}=-\left( 1-\frac{r^{2}}{l^{2}}\right) dt^{2}+\left( 1-\frac{r^{2}}{%
l^{2}}\right) ^{-1}dr^{2}+r^{2}d\Omega _{2}^{2}.
\end{equation} \label{FLRWS}

Also the FLRW metric (\ref{FLRWA}) could be put in the form

\begin{equation}
ds^{2}=-\frac{1-r^{2}/r_{A}^{2}}{1-kr^{2}/a^{2}}(dt+\frac{Hr}{%
1-r^{2}/r_{A}^{2}}dr)^{2}+\frac{1}{1-r^{2}/r_{A}^{2}}%
dr^{2}+r^{2}d\Omega _{2}^{2}. \label{ANTISITTER}
\end{equation} 

For this reason, Ya-Peng Hu observed that the de Sitter spacetime is only a particular case of the FLRW universe, in the previous equation $k=0$ and $r_{A}=H^{-1}=l$ are defined is a constant in equation (\ref{ANTISITTER}). Likewise, it is known that $r=l$ is the de Sitter space-time horizon, so the existence of a cosmological horizon in an FLRW universe is possible. By using the null cosmological horizon property and considering the spherical symmetry in equation (\ref{FLRWA}), the corresponding cosmological horizon $r=r_{H}(t)$ satisfies the constraint \cite{Ya-Peng}

\begin{equation}
g^{\mu \nu }\frac{\partial f}{\partial x^{\mu }}\frac{\partial f}{\partial
x^{\nu }}=0,
\end{equation} \label{constrict1}
is

\begin{equation}
1-r_{H}^{2}/r_{A}^{2}=\dot{r}_{H}^{2}-2Hr_{H}\dot{r}_{H}, \label{constrict2}
\end{equation} 
here $f=r-r_{H}(t)$. From equation (\ref{constrict2}), it was possible for Ya-Peng Hu to observe that the corresponding cosmological horizon $r_{H}(t)$ is just the cosmological horizon of the de Sitter space-time when $k=0$ and $\dot{r}_{H}=0$ \cite{Ya-Peng}.

Similarly, Ya Peng Hu considered the temperature of Hawking radiation coming from the cosmological horizon for $r=r_{H}(t)$ in an FLRW universe. For simplicity, in his work, the author considers the Klein-Gordon field in an FLRW universe. The Klein-Gordon equation in curved space-time is \cite{Ya-Peng}

\begin{equation}
(-m^{2})\psi =\frac{1}{\sqrt{-g}}\frac{\partial }{\partial x^{\mu }}(
\sqrt{-g}g^{\mu \nu }\frac{\partial }{\partial x^{v}})\psi -m^{2}\psi =0.
\end{equation}
In his paper, Ya-Peng Hu writes the Klein-Gordon equation (7) in terms of the scalar field $\Phi$. But later it will be necessary to talk about the wave function. For this reason, we will write the Klein-Gordon equation in terms of the wave function $\psi$. This equation can be rewritten in the FLRW coordinates (\ref{FLRWA}) such that \cite{Ya-Peng}

\begin{eqnarray}
& -\frac{\partial }{\partial t}(\frac{1}{\sqrt{1-\frac{k}{a^{2}}r^{2}}}\frac{%
\partial }{\partial t})\frac{\rho (t,r)}{r}-\frac{\partial }{\partial t}(%
\frac{Hr}{\sqrt{1-\frac{k}{a^{2}}r^{2}}}\frac{\partial }{\partial r})\frac{%
\rho (t,r)}{r}-\frac{1}{r^{2}}\frac{\partial }{\partial r}(\frac{r^{2}}{%
\sqrt{1-\frac{k}{a^{2}}r^{2}}}Hr\frac{\partial }{\partial t})\frac{\rho (t,r)%
}{r}  \nonumber \\ 
& +\frac{1}{r^{2}}\frac{\partial }{\partial r}[\frac{r^{2}}{\sqrt{1-\frac{k}{%
a^{2}}r^{2}}}(1-r^{2}/r_{A}^{2})\frac{\partial }{\partial r}]\frac{\rho (t,r)%
}{r}=[m^{2}+\frac{l(l+1)}{r^{2}}]\frac{1}{\sqrt{1-\frac{k}{a^{2}}r^{2}}}%
\frac{\rho (t,r)}{r} 
\end{eqnarray}
and

\begin{equation}
\frac{1}{\sin \theta }\frac{\partial }{\partial \theta }(\sin \theta \frac{%
\partial }{\partial \theta })Y_{lm}(\theta ,\varphi )+\frac{1}{\sin
^{2}\theta }\frac{\partial ^{2}}{\partial \varphi ^{2}}Y_{lm}(\theta
,\varphi )+l(l+1)Y_{lm}(\theta ,\varphi )=0,
\end{equation}
where $m$ is the rest mass of the Klein-Gordon particle, $Y_{lm}(\theta,\varphi )$ are the usual spherical harmonics and $\psi $ has been separated
as \cite{Ya-Peng}

\begin{equation}
\psi =\frac{1}{r}\rho (t,r)Y_{lm}(\theta,\varphi ).
\end{equation}

The scalar field near the cosmological horizon behaves according to the generalized comoving coordinate transformation \cite{Ya-Peng}

\begin{equation}
r_{\ast }=r+\frac{1}{2\kappa }\ln [r_{H}(t)-r]
\end{equation}  
and 

\begin{equation}
t_{\ast }=t-t_{0},    
\end{equation}
where $\kappa $ is the surface gravity, $r_{H}(t)$ is the location of the cosmological horizon, and $t_{0}$ is a time constant that describes the time at which particles radiate from the horizon. It is important to consider that $\kappa $ here is the surface gravity of the event horizon or cosmological horizon in stationary spacetime \cite{Ya-Peng}.

From (11) and (12), the radial equation (8) becomes \cite{Ya-Peng}

\begin{equation}
\begin{array}{ccc}
\{-\frac{2\kappa (r-r_{H})(\overset{.}{a}^{2}+k+a\overset{..}{a})}{
a[r(2r\kappa -2r_{H}\kappa +1)\overset{.}{a}-a\overset{.}{r}_{H}]}+\frac{
2\left( l^{2}+l+m^{2}r^{2}\right) \kappa a(r-r_{H})}{r^{2}[r(2r\kappa
-2r_{H}\kappa +1)\overset{.}{a}-a\overset{.}{r}_{H}]}\}\rho \\ \\
+\{\frac{-[\overset{.}{r}_{H}^{2}+(r-r_{H})\overset{..}{r}
_{H}-1]a^{2}+[(r+r_{H})\overset{.}{a}\overset{.}{r}_{H}+r(r-r_{H})(2r\kappa
-2r_{H}\kappa +1)\overset{..}{a}]a}{a(r-r_{H})[r(2r\kappa -2r_{H}\kappa +1)
\overset{.}{a}-a\overset{.}{r}_{H}]}
\\ \\ +\frac{r[2\kappa r^{2}+2\kappa r_{H}^{2}-(4r\kappa +1)r_{H}](\overset{.}{a}
^{2}+k)}{a(r-r_{H})[r(2r\kappa -2r_{H}\kappa +1)\overset{.}{a}-a\overset{.}{r
}_{H}]}\}\frac{\partial \rho }{\partial r_{\ast }}+\{\frac{(2r\kappa
-2r_{H}\kappa +1)^{2}(\overset{.}{a}^{2}+k)r^{2}}{2\kappa
a(r-r_{H})[r(2r\kappa -2r_{H}\kappa +1)\overset{.}{a}-a\overset{.}{r}_{H}]}
\\ \\ +\frac{-2(2r\kappa -2r_{H}\kappa +1)\overset{.}{a}\overset{.}{r}
_{H}ar+a^{2}[-(2r\kappa +1)^{2}+4\kappa r_{H}(2r\kappa +1)-4\kappa
^{2}r_{H}^{2}+\overset{.}{r}_{H}^{2}]}{2\kappa a(r-r_{H})[r(2r\kappa
-2r_{H}\kappa +1)\overset{.}{a}-a\overset{.}{r}_{H}]}\}\frac{\partial
^{2}\rho }{\partial r_{\ast }^{2}} 
\\ \\ +\frac{2\kappa (r-r_{H})\overset{.}{a}}{r(2r\kappa -2r_{H}\kappa +1)
\overset{.}{a}-a\overset{.}{r}_{H}}\frac{\partial \rho }{\partial t_{\ast }}+2\frac{\partial ^{2}\rho }{\partial t_{\ast }\partial r_{\ast }}+\frac{
2\kappa a(r-r_{H})}{r(2r\kappa -2r_{H}\kappa +1)\overset{.}{a}-a\overset{.}{r
}_{H}}\frac{\partial ^{2}\rho }{\partial t_{\ast }^{2}}=0.         
\end{array}
\end{equation}

When $r\rightarrow r_{H}$ and $t\rightarrow t_{0}$, the radial equation (13) is \cite{Ya-Peng}

\begin{equation}
A\frac{\partial ^{2}\rho }{\partial r_{\ast }^{2}}+2\frac{\partial ^{2}\rho
}{\partial t_{\ast }\partial r_{\ast }}+\alpha _{0}\frac{\partial \rho }{\partial r_{\ast }}=0.
\end{equation}

In \cite{Ya-Peng}, equation was used (13) and

\begin{equation}
A=-\frac{H\overset{.}{r}_{H}-(H^{2}+k/a^{2})r_{H}}{\kappa (Hr_{H}-\overset{.}%
{r}_{H})}+2\overset{.}{r}_{H},~\alpha _{0}=\frac{(H^{2}+k/a^{2})r_{H}-H%
\overset{.}{r}_{H}+\overset{..}{r}_{H}-\frac{\overset{..}{a}}{a}r_{H}}{%
\overset{.}{r}_{H}-Hr_{H}}.
\end{equation}

The two linearly independent solutions of (14) are \cite{Ya-Peng}

\begin{equation}
\rho _{out}=e^{-i\omega t_{\ast }},
\end{equation}
and

\begin{equation}
\rho _{in}=e^{-i\omega t_{\ast }+2i\omega r_{\ast }/A}e^{-\alpha
_{0}r_{\ast }/A}.
\end{equation}

These expressions represent the radial component for the scalar field within the cosmological horizon ($r<r_{H}$). The Klein-Gordon equation in comoving coordinates can be reduced to the standard form of the wave equation near the horizon \cite{Ya-Peng}, \cite{Damour_3}, \cite{Zhao}, \cite{Li_1} and \cite{Hu}.

\begin{equation}
\frac{\partial ^{2}\rho }{\partial r_{\ast }^{2}}+2\frac{\partial ^{2}\rho }{%
\partial t_{\ast }\partial r_{\ast }}=0.
\end{equation}

It is possible to adjust the parameter $\kappa$ to make $A=1$, and obtain \cite{Ya-Peng}

\begin{equation}
\kappa =\frac{H\overset{.}{r}_{H}-(H^{2}+k/a^{2})r_{H}}{(Hr_{H}-\overset{.}{r}_{H})(2\overset{.}{r}_{H}-1)}.
\end{equation}

Here it is necessary to take into account that $A=1$ can be used in the following special case for the de Sitter spacetime. Take into account that, $k=0$ and $\dot{r}_{H}=0$
with $r_{A}=H^{-1}=l$, the $\kappa$ in (19) is $\kappa=1/l$ which is simply the surface gravity of the cosmological horizon in the de Sitter spacetime. If equation (19) is taken into account, it is possible to find that $\kappa$ is in fact a constant related to the variable $t_{0}$ \cite{Ya-Peng}.

Therefore, the ingoing wave of the Klein-Gordon field near the cosmological horizon can be further rewritten as \cite{Ya-Peng}

\begin{equation}
\rho _{in}=Ce^{-i\omega t_{\ast }+2i\omega r_{\ast }}e^{-\alpha _{0}r_{\ast
}}=Ce^{-i\omega t_{\ast }}e^{2i\omega r-\alpha _{0}r}(r_{H}-r)^{i\omega
/\kappa -\alpha _{0}/2\kappa },
\end{equation}
where \cite{Ya-Peng} has used (11) and (12) and added the normalized factor $C$. Note that, $\rho _{out}$ represents an outgoing wave and is well-behaved when analytically extended outside $r>r_{H}$. However, we can find that the ingoing
wave $\rho _{in}$ (20) has a logarithmic singularity at the cosmological horizon $r=r_{H}$ and
is not analytical on the cosmological horizon. Thus
we can extend it by analytical continuation from the inside to the outside of the cosmological
horizon \cite{Ya-Peng}, \cite{Damour_3}, \cite{Zhao}, \cite{Li_1} and \cite{Hu}

\begin{equation}
(r_{H}-r)\rightarrow |r_{H}-r|e^{i\pi }=(r-r_{H})e^{i\pi }.
\end{equation}

Thanks to this, the incoming wave (20) becomes \cite{Ya-Peng}

\begin{equation}
\begin{array}{cc}
\rho _{in}\rightarrow
{\tilde \rho }_{in}=Ce^{-i\omega
t_{\ast }}e^{2i\omega r-\alpha _{0}r}(r-r_{H})^{i\omega /\kappa -\alpha
_{0}/2\kappa }e^{-\frac{i\pi \alpha _{0}}{2\kappa }}e^{-\frac{\pi \omega }{
\kappa }} & \\\\ =Ce^{-i\omega t_{\ast }+2i\omega r_{\ast }}e^{-\alpha _{0}r_{\ast
}}e^{-\frac{i\pi \alpha _{0}}{2\kappa }}e^{-\frac{\pi \omega }{\kappa }
},r>r_{H}.
\end{array}
\end{equation}

Using the Heaviside function $Y$, in \cite{Ya-Peng} it was found that

\begin{equation}
Y(x)=\Big\{_{0,~~~x~<0}^{1,~~~x~\geq 0}
\end{equation}
the complete ingoing wave can be \cite{Ya-Peng}

\begin{equation}
\phi _{\omega }^{in}=N_{\omega }[Y(r_{H}-r)\rho _{in}+Y(r_{H}-r){\tilde \rho }_{in}], 
\end{equation}
where $N_{w}$ is a normalization factor \cite{Ya-Peng}.

Here as in \cite{Ya-Peng}, we represent an incoming particle wave within the cosmological horizon and an outgoing antiparticle wave of negative energy outside the cosmological horizon (20) and (22) \cite{Damour_3}. It has also been mentioned previously that another interpretation is that an antiparticle of positive energy incoming in the past disperses forward in time on the cosmological horizon, and the incoming wave that describes this antiparticle state is simply the equation (24), \cite{Ya-Peng}.

It is possible to construct an analogy with the WKB approximation in quantum mechanics barrier penetration, $N_{\omega }^{2}$ can represent the force of a particle wave entering or tunneling from the cosmological horizon, \cite{Ya-Peng}.

The explanation is as follows, the state of the antiparticle $\phi _{\omega }^{in}$ is divided into two components. The first component is a force particle wave $N_{\omega }^{2}$ entering from the horizon and the second component is a negative antiparticle energy flow wave $N_{\omega }^{2} $ outgoing in the future towards the outside of the cosmological horizon. This represents an antiparticle force wave $N_{\omega }^{2}e^{-2 \pi \omega /\kappa}$ that has a flow of positive energy incoming in the past from outside the cosmological horizon \cite{Ya-Peng}, \cite{Damour_3}, \cite{Zhao}, \cite{Li_1} and \cite{Hu}.

We can use $\rho _{in}$, which is already normalized, and obtain the scalar product of $\phi _{\omega }^{in}$ in (24), in order to obtain the temperature of the Hawking radiation given in \cite{Ya-Peng}, \cite{Damour_3}, \cite{Zhao}, \cite{Li_1} and \cite{Hu}

\begin{equation}
(\phi _{\omega _{1}}^{in},\phi _{\omega _{2}}^{in})=N_{\omega _{1}}N_{\omega
_{2}}(\delta _{\omega _{1}\omega _{2}}-e^{-\pi (\omega _{1}+\omega
_{2})/\kappa }\delta _{\omega _{1}\omega _{2}}),
\end{equation}
here the inner product of the wave function was used. To obtain negative energy in equation (25) for bosonic particles, it is necessary to normalize the negative $\delta$ function. Now, taking into account that, if $\kappa <0$ in (25), it is possible to rewrite the inner product as  \cite{Ya-Peng}
\begin{equation}
(\phi _{w}^{in},\phi _{w}^{in})=-1=N_{\omega }^{2}(1-e^{-2\pi \omega /\kappa}).
\end{equation}

The latter can be associated with a thermal spectrum for a temperature $T=-\kappa /2\pi $. Likewise, if we consider that $\kappa >0$ we have (see \cite{Ya-Peng})

\begin{equation}
(\phi _{\omega }^{in},\phi _{\omega }^{in})=1=N_{\omega }^{2}(1-e^{-2\pi
\omega /\kappa }).
\end{equation}

Although strictly speaking it is not a thermal spectrum. In his work, Ya-Peng Hu was able to redefine the complete incoming wave as (24) in \cite{Ya-Peng}

\begin{equation}
\phi _{\omega }^{in^{\prime }}=e^{\frac{\pi \omega }{\kappa }}N_{\omega
}[Y(r_{H}-r)\rho _{in}+Y(r_{H}-r)\overset{\symbol{126}}{\rho }_{in}],
\end{equation}
from which we obtain \cite{Ya-Peng}

\begin{equation}
(\phi _{\omega }^{in^{\prime }},\phi _{\omega }^{in\prime })=1=N_{\omega
}^{2}(e^{2\pi \omega /\kappa }-1).
\end{equation}

This is a thermal spectrum with the temperature $T=\kappa /2\pi $ \cite{Ya-Peng}.

Thus the thermal spectrum was obtained in \cite{Ya-Peng} in both cases

\begin{equation}
N_{\omega }^{2}=\frac{1}{\exp (\omega /K_{B}T)-1},
\end{equation}
and the temperature $T$ is

\begin{equation}
T=\frac{|\kappa |}{2\pi }=|\frac{(H^{2}+k/a^{2})r_{H}-H\overset{.}{r}_{H}}{%
2\pi (Hr_{H}-\overset{.}{r}_{H})(2\overset{.}{r}_{H}-1)}|.
\end{equation}

The entropy can be obtained considering \cite{Hawking_1}, \cite{Hawking_2}, \cite{Hawking_3} and \cite{Hawking_4}

\begin{equation}
dQ=TdS=\frac{\kappa}{8 \pi} dA,    
\end{equation}
and it is possible to understand that entropy can be expressed in the following way

\begin{equation}
S=\frac{\kappa}{8 \pi} \frac{A}{T}.   
\end{equation}

This entropy can be expressed in terms of the temperature (31) as

\begin{equation}
S=\frac{\kappa}{8 \pi T} \frac{2\pi }{|\kappa |}=|\frac{\kappa}{8 \pi} \frac{2\pi (Hr_{H}-\overset{.}{r}_{H})(2\overset{.}{r}_{H}-1)}{(H^{2}+k/a^{2})r_{H}-H\overset{.}{r}_{H}}|.
\end{equation}

The entropy $S$ is related to the number of bits $N$ by the equation

\begin{equation}
S=\frac{\kappa N l_{p} ^{2}}{4 \pi},    
\end{equation}
here $l_{p}$ is the Planck length and the Planck area is expressed as $l_{p}^2=(\hbar G)/(c^{3})$. Also the number of bits is expressed as 

\begin{equation}
N=\frac{1}{2}| \frac{2\pi (Hr_{H}-\overset{.}{r}_{H})(2\overset{.}{r}_{H}-1)}{(H^{2}+k/a^{2})r_{H}-H\overset{.}{r}_{H}}| \frac{A}{l_{p} ^{2}}.
\end{equation}

From here it can be seen that the entropy of the universe is a function of the number of bits. This is known in some physical scenarios as a holographic principle.

It can be seen, that there is a way to express the entropy of the universe according to classical information theory.

In this sense, it is considered that the temperature of the universe contributes to its entropy and consequently to the amount of information that can be contained within the surface of the universe. Thus the information is stored on the two-dimensional boundary of the universe and it is projected in three dimensions, which gives shape to matter. One of the most studied aspects of this principle is the AdS/CFT correspondence, in which it is considered that the physics modeled on a 4-dimensional curved surface for an anti-de Sitter spacetime (solution of Einstein's equations with negative cosmological constant). This presents an equivalence with conformal field theories (CFT) which are quantum field theories defined on a boundary with a dimension smaller than the AdS theory \cite{Maldacena}.

de Sitter spacetime may be a particular case of an FLRW universe; in the cosmological horizon of de Sitter spacetime, there is Hawking radiation, this was one of the conclusions of Ya-Peng Hu in \cite{Ya-Peng}.

Ya-Peng Hu mentions that “there are some clues that show that there is Hawking radiation at the apparent horizon in an FLRW universe” \cite{Ya-Peng}. This was done by understanding how a Klein-Gordon field behaves near the cosmological horizon. With this, it was understood that Hawking radiation comes from the cosmological horizon of an FLRW Universe. It is important to emphasize here that, when $\dot{r}_{H}=0$, the cosmological horizon at (6) is the same as the apparent horizon, and the temperature in (31) is \cite{Ya-Peng}

\begin{equation}
T=\frac{1}{2\pi Hr_{A}^{2}}.
\end{equation}

The previous temperature differs from the expression for the Hawking radiation that other authors have previously obtained for the apparent horizon $T=1/(2\pi r_{A})$ \cite{Cai}, \cite{Li_2}. As it can be seen, this expression does not incorporate the Hubble parameter; intending to obtain a more coherent cosmological model, this parameter must be considered. At that point the temperature $T=1/(2\pi r_{A})$ is measured by a Kodama-type observer \cite{Ya-Peng}.

This Kodama-type observer is based on thermodynamic variables defined by a Killing observer, which is absent in the time-dependent case, since our universe is globally a non-static solution of the gravitational field equations. The thermodynamic analogy of gravity in time-dependent spacetime goes deeper. This difficulty could be somewhat solved by a geometrically natural representation: a divergence-free preferred vector field proposed by Kodama \cite{Kodama}. The so-called “Kodama vector” defines a natural time direction in the general spherically symmetric spacetime, and induces an unexpected conserved charge, which coincides with the Misner energy \cite{Ya-Peng}, \cite{Kodama} and \cite{Yi-Xing}.

The temperature measured by the observer

$$
(\partial / \partial t)^{a} \nonumber
$$
in equation (\ref{FLRWA}) is $T=1/(2\pi Hr_{A}^{ 2})$ \cite{Cai}. Consider that the radial velocity $\dot{r}_{H}=0$, ensures the observer in the coordinate system (11) and (12) is the same as the observer $(\partial / \partial t)^{a} $ in (\ref{FLRWA}) \cite{Ya-Peng}, \cite{Kodama} and \cite{Yi-Xing}.

This condition $\dot{r}_{H}=0$ is consistent with the underlying integrability condition in \cite{Cai} and \cite{Li_2}. From $\dot{r}_{H}=0$, it was possible in \cite{Ya-Peng} to find that the cosmological horizon and the apparent horizon are equal. His conclusion was that $\dot{R}_{H} = 0$ can be reduced to $\dot{R}_{A} = 0$ (\cite{Ya-Peng} and \cite{Yi-Xing}).

The author also mentions that from equations (7) and (10) in \cite{Cai}, the underlying integrability condition comes from $\partial_{\tilde r} \partial_{t} S=\partial_{t} \partial_{\tilde r}S$ for this reason it can also be deduced that $\dot{r}_{A}=0$. With this, the author notes that his equation for temperature (31) is consistent with \cite{Ya-Peng} and \cite{Yi-Xing}.

\section{Variation of the Fundamental Constants with the Expansion of the Universe}

In particle physics, quantum electrodynamics (QED) is the relativistic theory of quantum electromagnetic fields. This describes how light interacts with matter. Quantum electrodynamics mathematically describes all phenomena that involve electrically charged particles through the exchange of photons.

The action for quantum electrodynamics is the following \cite{Peskin}

\begin{equation}
{S_{\text{QED}}=\int d^{4}x\,\left[-{\frac {1}{4}}F^{\mu \nu }F_{\mu \nu }+{\bar {\psi }}\,(i\gamma ^{\mu }D_{\mu }-m)\,\psi \right]},    
\end{equation}
where $\gamma ^{\mu }$ are Dirac matrices, $\psi$ a bispinor field of spin 1/2 particles, ${\bar {\psi }}\equiv \psi ^{\dagger }\gamma ^{0}$, called psi-bar, is sometimes referred to as the Dirac adjoint \cite{Peskin}.

The gauge covariant derivative is \cite{Peskin}

\begin{equation}
{\displaystyle D_{\mu }\equiv \partial _{\mu }+ieA_{\mu }+ieB_{\mu }},    
\end{equation}
where $e$ is the coupling constant for the electric charge, equal to the electric charge of the bispinor field, $A_{\mu }$ is the 4-covariant potential of the electromagnetic field coming from the electron. This field have the symmetry group ${\displaystyle {\text{U}}(1)}$, $B_{\mu }$ is the magnetic field coming from an external source and $m$ is the mass of the electron or positron \cite{Peskin}. 

Now, the electromagnetic field tensor is \cite{Peskin}

\begin{equation}
{\displaystyle F_{\mu \nu }=\partial _{\mu }A_{\nu }-\partial _{\nu }A_{\mu }}. 
\end{equation}

The covariant derivative is useful when developing a Lagrangian \cite{Peskin}

 \begin{equation}
{\displaystyle {\mathcal {L}}=-{\frac {1}{4}}F_{\mu \nu }F^{\mu \nu }+{\bar {\psi }}(i\gamma ^{\mu }\partial _{\mu }-m)\psi -ej^{\mu }A_{\mu }},     
 \end{equation}
where $j^{\mu }$ is the conserved current under the action of the ${\displaystyle {\text{U}}(1)}$ group arising from Noether's theorem. It is written ${\displaystyle j^{\mu }={\bar {\psi }}\gamma ^{\mu }\psi}$ \cite{Peskin}.

The covariant derivative can now be expanded in the Lagrangian to obtain \cite{Peskin}

\begin{align} 
\begin{array}{cc}
& \mathcal {L}  = -{\frac{1}{4}}F_{\mu \nu} F^{\mu \nu}+i{\bar {\psi}}\gamma^{\mu} \partial _{\mu} \psi -e{\bar {\psi}}\gamma^{\mu}A_{\mu}\psi -m{\bar {\psi}}\psi \\ 
\\
&= -{\frac{1}{4}} F_{\mu \nu}F^{\mu \nu}+i{\bar {\psi}}\gamma^{\mu}\partial _{\mu}\psi -m{\bar {\psi}}\psi -ej^{\mu}A_{\mu}.
\end{array}    
\end{align} 

The external magnetic field $B_{\mu }$ has been set to zero, for simplicity. Alternatively, we can absorb $B_{\mu }$ into a new gauge field ${\displaystyle A'_{\mu }=A_{\mu }+B_{\mu }}$ and relabel the new field as ${\displaystyle A_{\mu }}$ \cite{Peskin}.

By developing the Euler-Lagrange equations for this Lagrangian and for ${\displaystyle {\bar {\psi }}}$, the equations of motion for the fields $\psi$ and $A_{\mu }$ can be obtained \cite{Peskin}.

This Lagrangian does not contain terms ${\displaystyle \partial _{\mu }{\bar {\psi }}}$, so only the following term will be present \cite{Peskin}

\begin{equation}
{\displaystyle {\frac {\partial {\mathcal {L}}}{\partial {\bar {\psi }}}}=0},
\end{equation}
so the equation of motion can be written \cite{Peskin}

\begin{equation}
{\displaystyle (i\gamma ^{\mu }\partial _{\mu }-m)\psi =e\gamma ^{\mu }A_{\mu }\psi}.    
\end{equation}

Using the Euler–Lagrange equation for the $A_{\mu }$ field \cite{Peskin},

\begin{equation}
{\displaystyle \partial _{\nu }\left({\frac {\partial {\mathcal {L}}}{\partial (\partial _{\nu }A_{\mu })}}\right)-{\frac {\partial {\mathcal {L}}}{\partial A_{\mu }}}=0,}    
\end{equation}
the derivatives this time are \cite{Peskin}

\begin{equation}
{\displaystyle \partial _{\nu }\left({\frac {\partial {\mathcal {L}}}{\partial (\partial _{\nu }A_{\mu })}}\right)=\partial _{\nu }\left(\partial ^{\mu }A^{\nu }-\partial ^{\nu }A^{\mu }\right),}    
\end{equation}
and 

\begin{equation}
{\displaystyle {\frac {\partial {\mathcal {L}}}{\partial A_{\mu }}}=-e{\bar {\psi }}\gamma ^{\mu }\psi}.    
\end{equation}

The above can be substituted into equation (42) to have the following result \cite{Peskin}

\begin{equation}
{\displaystyle \partial _{\mu }F^{\mu \nu }=e{\bar {\psi }}\gamma ^{\nu }\psi },    
\end{equation}
which can be written in terms of the ${\displaystyle {\text{U}}(1)}$ current 
$j^{\mu }$ as in \cite{Peskin}

\begin{equation}
{\displaystyle \partial _{\mu }F^{\mu \nu }=ej^{\nu }.}    
\end{equation}

Here it is possible to impose a Lorentz gauge condition ${\displaystyle \partial _{\mu }A^{\mu }=0}$, and with this, the equation reduces to ${A^{\mu }=ej^{\mu }}$, which is a wave equation for the 4-potential \cite{Peskin}.

Now that all of these QED considerations have been made, and knowing that we are working with the FLRW metric, it is more convenient to write the 4-electromagnetic momentum in a curved spacetime:

\begin{equation}
P_\mu=(E/c,P)=\hbar K^{\mu}= \hbar (\omega/c, \vec{k}),
\end{equation}
thus the 4-wavevector is related to the 4-momentum \cite{Rindler}. In this equation the energy belonging to the electromagnetic field is

\begin{equation}
E=\int d^3 \vec{k} |\vec{k}| \sqrt g,    
\end{equation}
here $\vec{k}$ is the electromagnetic wave vector.

With this, the electromagnetic linear momentum can be calculated using:

\begin{equation}
P_\mu = \int d^3 K_\mu |\vec{k}| \sqrt g,   \end{equation}
where $K \mu$ is the electromagnetic wave vector component.

Energy and momentum are related by the energy-momentum equation; in a curved spacetime:

\begin{equation}
P_\mu P_\mu = -m^2 g_{\mu\nu}.     
\end{equation}

In spherical coordinates, the 4-electromagnetic momentum can be written as:

\begin{equation}
P_\mu = | E p\theta p\phi |,
\end{equation}
where $E=-iK_0$, $P_\theta = -iK_1$ and $P_\phi = -iK_2$. It should be taken into account that $K_0$ is the temporal component, $K_1$ the radial component and $K_2$ the angular component.

Consider the metric for a photon traveling in a gravitational field in spherical coordinates ($x=r \, sin \theta \, cos \phi$, $y=r \, sin \theta \, sin \phi$ and $z= r \, cos \theta$):

\begin{equation}
g_{\mu\nu}=\left( 
\begin{array}{ccc}
1 & 0 & 0 \\ 
0 & r^2 & 0 \\
0 & 0 & r^2 sin^2 \theta
\end{array}
\right).
\end{equation}

In curved space-time we have the quantities $E=-iK_0/\sqrt{1}$, $P_\theta = -iK_1/\sqrt{r^2}$ and $P_\phi = -iK_2 /\sqrt{r^2 sin^2 \theta}$.

The correct approximation is as the Ward-Takahashi equation for quantum field theory in curved spacetime, in which the 4-momentum can be written in the form \cite{Ward}, \cite{Takahashi} and \cite{Peskin}:

\begin{equation}
P_\mu = -i \int d^3 \vec{k} K_\mu |\vec{k}| \sqrt{g} \langle \psi |K| \psi \rangle.
\end{equation}

In order to describe the scalar field introduced in the previous sections, the Klein-Gordon equation coupled to the electromagnetic field will be used:

\begin{equation}
\frac{1}{\sqrt{-g}}\frac{\partial }{\partial x^{\mu }}(
\sqrt{-g}g^{\mu \nu }\frac{\partial }{\partial x^{v}})\psi -m^{2}\psi =e A_\mu \frac{1}{\sqrt{-g}} \frac{\partial }{\partial x^{\mu }} \psi.
\end{equation}

The wave function described in the above equation is used in spherical coordinates 

\begin{equation}
\psi =\frac{1}{r}\rho (t,r)Y_{lm}(\theta,\varphi ). \nonumber  
\end{equation}

The Klein-Gordon equation (57) can be rewritten as follows:

\begin{equation}
\begin{array}{cc}
-\frac{\partial }{\partial t}(\frac{1}{\sqrt{1-\frac{k}{a^{2}}r^{2}}}\frac{\partial}{\partial t})\frac{\rho(t,r)}{r}-\frac{\partial}{\partial t}(\frac{Hr}{\sqrt{1-\frac{k}{a^{2}}r^{2}}}\frac{\partial }{\partial r})\frac{\rho(t,r)}{r} \\ \\ -\frac{1}{r^{2}}\frac{\partial}{\partial r}(\frac{r^{2}}{\sqrt{1-\frac{k}{a^{2}}r^{2}}}Hr\frac{\partial}{\partial t})\frac{\rho(t,r)}{r} \\ \\ +\frac{1}{r^{2}}\frac{\partial }{\partial r}[\frac{r^{2}}{\sqrt{1-\frac{k}{a^{2}}r^{2}}}(1-r^{2}/r_{A}^{2})\frac{\partial }{\partial r}]\frac{\rho (t,r)}{r} \\ \\ +\frac{1}{\sin \theta }\frac{\partial }{\partial \theta }(\sin \theta \frac{\partial }{\partial \theta })Y_{lm}(\theta ,\varphi )+\frac{1}{\sin^{2}\theta }\frac{\partial ^{2}}{\partial \varphi ^{2}}Y_{lm}(\theta,\varphi ) \\ \\ =e A_r \frac{1}{\sqrt{-g}} \frac{1}{r^2} \frac{\partial }{\partial r} \rho(t,r)+e A_\theta \frac{1}{\sqrt{-g}} \frac{1}{\sin \theta } \frac{\partial }{\partial {\theta}} \rho(t,r) Y_{lm}(\theta,\varphi ) \\ \\ +e A_\varphi \frac{1}{\sqrt{-g}} \frac{1}{\sin^{2}\theta } \frac{\partial^2 }{\partial {\varphi^2}} \rho(t,r) Y_{lm}(\theta,\varphi ).
\end{array}
\end{equation}

The 4-electromagnetic momentum in curved spacetime (55) can be obtained from the Klein-Gordon equation coupled to Maxwell's equations, according to the following relation:

\begin{equation}
\langle \psi (\vec{k}) |A_\mu \partial_\mu | \psi(\vec{k}) \rangle =-i \int d^3 \vec{k} |\vec{k}| \sqrt{g} \langle \psi (\vec{k}) | \psi(\vec{k}) \rangle.     
\end{equation}

The above equation can be solved using the metric tensor given in (57). The photon wave function on which the integral (57) must be evaluated is 

\begin{equation}
\psi(\vec{k}) = e^{-i \vec{k} \dot \vec{x}},   
\end{equation}
which must be solved numerically. 

Furthermore, because we want to work within the context of the hypothesis of large Dirac numbers, which is proposed in the classical framework, to obtain the 4-momentum value of the electromagnetic field of the universe, semiclassical considerations must be made. In this work, we will limit ourselves to mentioning the equation (54) and pointing out its importance for finding relevant physical properties in cosmology that come from the presence of the scalar field described by the Klein-Gordon equation. In a curved space-time, the 4-momentum is written as:

\begin{equation}
P^\mu = \int ( \int F^{\mu} _{\nu} J^{\nu} \sqrt{-g} dV) d \tau, 
\end{equation}
in particular symmetric situations, this equation can be solved numerically.

The metric of the universe is the FLRW metric given in equation (\ref{FLRW}) and is written in spherical coordinates. With this, we have the following Klein-Gordon equation: the 4-momentum in spherical coordinates can be written as:

\begin{equation}
P^ \mu = \int d^3 x \sqrt{-g} T^{0 \nu}.
\end{equation}

It is also possible to couple the potential and electromagnetic field to the Klein-Gordon equation for a scalar field, this is reflected in the following expression for the action:

\begin{equation}
S= -\int d^4 x \sqrt{-g} [\frac{1}{2} \Box_\mu \phi \Box^{\mu} \phi -V(\phi) - \frac{1}{4} F_{\mu \nu} F^{\mu \nu} -J^{\mu} A_{\mu}].    
\end{equation}

With this, we have the following Klein-Gordon equation:

\begin{equation}
\Box_\mu \Box^{\mu} \phi - {\partial x^{\mu }V(\phi)} = J^{\mu} A_{\mu}.
\end{equation}

To simplify the calculations due to the mathematical complexity of expressions (55), (58), (59) and (60) for the 4-momentum, it is possible to use the integral:

\begin{equation}
P^\mu = \int A^\mu J^\mu dV,    
\end{equation}
where $dV=\int \Phi c \rho r^2 sin \theta dr d\theta d\phi$.

For cosmological considerations, a relationship between the 4-momentum, the Hubble parameter $H=\dot{a}(t)/(1+z)$ and the redshift $z$ can be obtained. The 4-momentum in special relativity is $E^{2}=p^{2}c^{2}+m^{2}c^{4}$, and it is also possible to consider Hubble's law as $v=H*d$ (here $[H]=[1/t]$), where the velocity is $v=c*z$.  Taking into account the above, the 4-momentum for the expansion of the universe as a function of the redshift $z$ can be written as:

\begin{equation}
P^{\mu}= (mc, m c^2 H \frac{z}{c} t, m c^2 H \frac{z}{c} t, m c^2 H \frac{z}{c} t).
\end{equation}

Another equation that relates the Hubble constant and the 4-electromagnetic momentum is the following (for dimensional consistency, all equations used to express the Hubble parameter will be in terms of Planck units):

\begin{equation}
H=\sqrt{(\frac{8 \pi G \rho}{3})}.
\end{equation}

These Planck units are the following: $t_p$ is the Planck time, $l_p$ is the Planck length, $\rho_p$ is the Planck density, $I_p$ is the Planck electrical intensity, $q_p$ is the Planck charge, $m_p$ is the Planck mass and $T_p$ is the Planck temperature. Taking into account the energy-momentum tensor for the energy of the
electromagnetic field \cite{Peskin}

\begin{equation}
T_{\sigma }^{\nu }=\frac{\partial \mathcal{L}_{m}}{\partial A_{\mu ,\nu }}%
A_{\mu ,\sigma }-\delta _{\sigma }^{\nu }\mathcal{L}_{m}. 
\end{equation}

With equation (68) it is possible to write this energy-momentum tensor for the propagation of the electromagnetic field \cite{Peskin}

\begin{equation}
\frac{1}{c}\int \left( T_{\sigma }^{\nu }+\delta _{\sigma }^{\nu }\mathcal{L}_{m}\right) dx^{\sigma }=0,
\end{equation}
using the Lagrangian density for the energy of the electromagnetic field 

\begin{equation}
\mathcal{L}_{m}=\frac{1}{8\pi }(E^{2}-H^{2}),    
\end{equation}
it is possible to find the energy density for the electromagnetic field \cite{Peskin}

\begin{equation}
T^{00}=\frac{1}{8\pi }(E^{2}+H^{2}).
\end{equation}

With this, it is possible to define the 4-momentum for the electromagnetic field as \cite{Peskin}

\begin{equation}
P^{\mu}=\int  T^{0 \nu} dV.   
\end{equation}

So by considering that the energy density $\rho=T^{0 \nu}/V$ it is possible to express the Hubble parameter (67) as

\begin{equation}
H=\sqrt{\frac{8 \pi G P^{\mu}}{3 V}}.
\end{equation}

This is consistent with what is seen in equation (66) in the following way

\begin{equation}
H=\sqrt{\frac{8 \pi G m c z t v \rho_{p} l_p {t_p}^2}{3 V D}}.
\end{equation}

In the previous equation, it was taken into account that the Hubble parameter can be expressed as $H=v/D$. Because we are interested in understanding what happens with the expansion of the universe, that is, when the radius $r_U$ of the universe varies, we use the radial component of the 4-momentum equation (63):

\begin{equation}
P^{1} = \int A^{r_U} \rho c r^2_U sin \theta dr_U d\theta d\phi.    
\end{equation}

This gives the following expression for the Hubble parameter $H$:

\begin{equation}
H=\frac{1}{c}\sqrt{\frac{8 \pi G {t_p}^4 l_p \int A^{r_U} \rho c r^2_U sin \theta dr_U d\theta d\phi}{3 V {l_p}^3 t_p m_p}}.
\end{equation}

Now that we have calculated the expression for the Hubble parameter (74), it is possible to find an expression for the temperature of the universe. We will consider the expression for Hawking radiation given in equation (31) \cite{Ya-Peng}

\begin{equation}
T=\frac{|\kappa |}{2\pi }=|\frac{(H^{2}+k/a^{2})r_{H}-H\overset{.}{r}_{H}}{2\pi (Hr_{H}-\overset{.}{r}_{H})(2\overset{.}{r}_{H}-1)}|.    
\end{equation}

Since we want to find the expression of the temperature for the universe and thereby understand how the fundamental constants such as the fine structure constant $\alpha=e^2/(4 \pi \epsilon_{0} \hbar c)$ can vary with the expansion of the entire universe, the cosmological parameters found in the expression (74), which are the redshift $z$, the radius of the universe expressed as in the Dirac large numbers hypothesis $(r_U/r_e) = (r_H/r_e)$. As it can be seen in the Dirac large numbers hypothesis, the classical radius of the electron $r_e = e^2/(4 \pi \epsilon _0 m_e c^2)$ appears with an electron mass $m_e$, an electrostatic radius $r_H = e^2/(4 \pi \epsilon _0 m_e c^2)$, with a mass for a hypothetical particle $m_H=(G m_e^2)/(r_e c^2)$. 

Because we are also interested in the holographic principle, the number of bits given for a spherical surface area will be considered $N= A/(l^2_p)= (c^3 A)/(G \hbar)$, the critical density of the universe $\rho_c = (3 H^2)/(8 \pi G)$, $\rho(a)=\rho_{0}/a^{4}$ and the proportionality constant of the Hubble parameter $H_0 =(zc)/(r_U)$ \cite{Dirac_1}, \cite{Dirac_2} and \cite{Dirac_3}.

If the charge distribution is approximately spherical but inhomogeneous, we have:

\begin{equation}
A^{r_U} = \frac{1}{4 \pi \epsilon_0} \int _{0}^{2 \pi} \int _{0}^{\pi} \int _{0}^{R_U} \frac{r'^2_U \rho(r',\theta, \phi)}{r'_U} sin \theta dr_U' d \theta d \phi = \frac{{l_p}^2 q_p R^3 _U \rho_0}{3 \epsilon_0 m_p {t_p}^4}.
\end{equation}

Due to the above, it is possible to solve the integral in the equation (76) and thereby find a value for the temperature of the universe from the Hawking radiation equation. To solve the equation (78) it is necessary to substitute the value of the equation (75) given in the equation for the Hubble parameter (76). Explicitly the solution to equation (76) is

\begin{equation}
H=\sqrt{\frac{32 \pi^{2}G\rho_U\rho_0 {t_p}^2}{9\epsilon_{0} V {l_p} {q_p}^2}\frac{r_U ^6}{6}}.
\end{equation}

Here it is necessary to consider that the critical density of the universe is $\rho_U=\rho_c=3H^2/8 \pi G$, where the number of bits $N$ has been taken into account.

By substituting the Hubble parameter from equation (79) into the Hawking radiation equation (77), a complicated integral is obtained. To simplify this, an approximation is made using a Taylor expansion for the denominator of equation (77), where the small parameter is $H$ as in equation (79):

\begin{equation}
\begin{array}{ccc}
T=\frac{|\kappa |}{2\pi }=|({(H^{2}+k/a^{2})r_{H}-H\overset{.}{r}_{H}})*({\frac{1}{2\pi \overset{.}{r}_{H} - 4\pi \overset{.}{r}^2_{H}}} && \\ +{\frac{r_{H} H}{2\pi \overset{.}{r}^2_{H} - 4\pi \overset{.}{r}^3_{H} }}+{\frac{{r^2}_{H} H^2}{2\pi \overset{.}{r}^3_{H} - 4\pi \overset{.}{r}^4_{H} }}+{\frac{{r^3}_{H} H^3}{2\pi \overset{.}{r}^4_{H} - 4\pi \overset{.}{r}^5_{H} }}+...)|.  
\end{array}
\end{equation}

By truncating this expansion to the first order, from equation (80) we can solve for the Hubble parameter (For $r/\overset{.}{r}_{H} \neq 0$ and $T=0$):

\begin{equation}
H=\frac{\pi \overset{.}{r}_{H}({-\frac{1}{2 \pi} \pm \sqrt{\frac{1}{4 \pi^2}-\frac{k {r^2}_{H}}{\pi^2 a^2 {t_p}^2 \overset{.}{r^2}_{H}}}})}{r_{H}}.  
\end{equation}

Likewise, it is now possible to truncate this expansion to second order and obtain the following equations for the Hubble parameter:

a) With $r_{H}=0$, the constraint $2 a^2 \overset{.}{r}^2_{H} - a \overset{.}{r}_{H} \neq 0$ and $T=0$

\begin{equation}
H=0.    
\end{equation}

b) With $r_{H}=0$, the constraint $2 a^2 \overset{.}{r}^2_{H} - a \overset{.}{r}_{H} \neq 0$ and $T=0$

\begin{equation}
H=\frac{1}{t_p} \frac{(\frac{t_p}{l_p}) \overset{.}{r}_{H} - 2 (\frac{{t_p}^2}{{l_p}^2}) \overset{.}{r}^2_{H}}{\frac{t_p}{l_p} \overset{.}{r}_{H}}.    
\end{equation}

c) With $a r_{H} \neq 0$, the constraint $2 \overset{.}{r}^2_{H} - \overset{.}{r}_{H} \neq 0$ and $T=0$

\begin{equation}
H=\frac{l_p}{t_p} \frac{a^2 (\frac{t_p}{l_p}) \overset{.}{r}_{H} - \sqrt{a^2 (a^2 (\frac{{t_p}^2}{{l_p}^2}) \overset{.}{r}^2_{H} -4k (\frac{{t_p}^2}{{l_p}^2}) \overset{.}{r}^2_{H} )}}{2 a^2 r_{H}}.    
\end{equation}

d) With $a r_{H} \neq 0$, $2 \overset{.}{r}^2_{H} - \overset{.}{r}_{H} \neq 0$ and $T=0$

\begin{equation}
H=\frac{l_p}{t_p} \frac{a^2 (\frac{{t_p}}{{l_p}})  \overset{.}{r}_{H} + (a^2 (a^2 (\frac{{t_p}^2}{{l_p}^2}) \overset{.}{r}^2_{H} -4k (\frac{{t_p}^2}{{l_p}^2}) \overset{.}{r}^2_{H} ))}{2 a^2 r_{H}}.     
\end{equation}   

To use the expression for the Taylor series expansion to the first order the Hawking radiation for temperature (77), it is necessary to consider that $T \neq 0$, with this we have the following cases:

a) $r_{H} = 0$ and $a \overset{.}{r}_{H} T \neq 0$, then we obtain of the following

\begin{equation}
H= \frac{2 \pi T}{{t_p} {T_p}}.    
\end{equation}

b) $a r_{H} = 0$ and $a \overset{.}{r}_{H} T \neq 0$, then we obtain of the following

\begin{equation}
H= \frac{1}{t_p} \frac{a^2 (\frac{{t_p}}{{l_p}}) \overset{.}{r}_{H} - \sqrt{a^2 (-8 \pi a^2 (\frac{{t_p}}{{l_p}}) \frac{r_{H}}{l_p} \overset{.}{r}_{H} \frac{T}{T_p} + a^2 (\frac{{t_p}^2}{{l_p}^2}) \overset{.}{r}^2_{H}-4k (\frac{{t_p}^2}{{l_p}^2}) \overset{.}{r}^2_{H}})}{2 a^2 {l_p} \overset{.}r_{H}}.    
\end{equation}

c) $a r_{H} = 0$ and $a \overset{.}{r}_{H} T \neq 0$, then we have the following

\begin{equation}
H= \frac{1}{t_p} \frac{{( \sqrt{a^2 (-8 \pi a^2 (\frac{{t_p}}{{l_p}}) \frac{r_{H}}{l_p} \overset{.}{r}_{H} \frac{T}{T_p} + a^2 (\frac{{t_p}^2}{{l_p}^2}) \overset{.}{r}^2_{H}-4k (\frac{{t_p}^2}{{l_p}^2}) \overset{.}{r}^2_{H}}+a^2 (\frac{{t_p}}{{l_p}}) \overset{.}{r_{H}}}}{2 a^2 \overset{.}{r}_{H}}.  
\end{equation}

At this point, the second-order expansion of the Hawking radiation is not considered since it is no longer possible to find an exact mathematical solution from the resulting equation for the Hubble parameter.

Considering that the scale parameter can be expressed as $a=1/(1+z)$ it is possible to reexpress the previous equations b) and c) in terms of the redshift $z$:

a) $a r_{H} = 0$ and $a \overset{.}{r}_{H} T \neq 0$, we have

\begin{equation}
\begin{array}{cc}
H= \frac{1}{t_p} \frac{(\frac{1}{1+z})^2 (\frac{{t_p}}{{l_p}}) \overset{.}{r}_{H} - \sqrt{(\frac{1}{1+z})^2 (-8 \pi (\frac{1}{1+z})^2 (\frac{{t_p}}{{l_p}}) \frac{r_{H}}{l_p} \overset{.}{r}_{H} \frac{T}{T_p} + (\frac{1}{1+z})^2 (\frac{{t_p}^2}{{l_p}^2}) \overset{.}{r}^2_{H}-4k (\frac{{t_p}^2}{{l_p}^2}) \overset{.}{r}^2_{H}})}{2 (\frac{1}{1+z})^2 {l_p} \overset{.}r_{H}}.    
\end{array}
\end{equation}

b) $a r_{H} = 0$ and $a \overset{.}{r}_{H} T \neq 0$, the following holds

\begin{equation}
\begin{array}{cc}
H= \frac{1}{t_p} \frac{{( \sqrt{(\frac{1}{1+z})^2 (-8 \pi (\frac{1}{1+z})^2 (\frac{{t_p}}{{l_p}}) \frac{r_{H}}{l_p} \overset{.}{r}_{H} \frac{T}{T_p} + (\frac{1}{1+z})^2 (\frac{{t_p}^2}{{l_p}^2}) \overset{.}{r}^2_{H}-4k (\frac{{t_p}^2}{{l_p}^2}) \overset{.}{r}^2_{H}}+(\frac{1}{1+z})^2 (\frac{{t_p}}{{l_p}}) \overset{.}{r_{H}}}}{2 (\frac{1}{1+z})^2 \overset{.}{r}_{H}}.  
\end{array}
\end{equation}

Now to use the hypothesis of Large Dirac Numbers it is necessary to replace the expression of the Hubble parameter (79) in the expressions (89) and (90). With this, it is possible to express the radius of the universe in the form $r_U= e^2/(4 \pi \epsilon _0 m_e c^2)=(\alpha \hbar)/(m_{e} c)$, that $\rho_U=\rho_c = (3 H^2)/(8 \pi G)$ and knowing that $H=\overset{.}{a(t)}/a(t)$: 

a) $a r_{H} = 0$ and $a \overset{.}{r}_{H} T \neq 0$, we have

\begin{equation}
\begin{array}{cc}
H= \frac{1}{t_p} \sqrt{(\frac{432 c^4 {m_{e}}^6 \epsilon_{0} \hbar G N}{96 \pi {m_p}^5 l_p A_{H} \rho_0 \alpha^6 \hbar^6})} \frac{(\frac{1}{1+z})^2 (\frac{{t_p}}{{l_p}}) \overset{.}{r}_{H} - \sqrt{(\frac{1}{1+z})^2 (-8 \pi (\frac{1}{1+z})^2 (\frac{{t_p}}{{l_p}}) \frac{r_{H}}{l_p} \overset{.}{r}_{H} \frac{T}{T_p} + (\frac{1}{1+z})^2 (\frac{{t_p}^2}{{l_p}^2}) \overset{.}{r}^2_{H}-4k (\frac{{t_p}^2}{{l_p}^2}) \overset{.}{r}^2_{H}})}{2 (\frac{1}{1+z})^2 {l_p} \overset{.}r_{H}}. 
\end{array}
\end{equation}

b) $a r_{H} = 0$ and $a \overset{.}{r}_{H} T \neq 0$ is obtained

\begin{equation}
\begin{array}{cc}
H= \frac{1}{t_p} \sqrt{(\frac{432 c^4 {m_{e}}^6 \epsilon_{0} \hbar G N}{96 \pi {m_p}^5 l_p A_{H} \rho_0 \alpha^6 \hbar^6})} \frac{{( \sqrt{(\frac{1}{1+z})^2 (-8 \pi (\frac{1}{1+z})^2 (\frac{{t_p}}{{l_p}}) \frac{r_{H}}{l_p} \overset{.}{r}_{H} \frac{T}{T_p} + (\frac{1}{1+z})^2 (\frac{{t_p}^2}{{l_p}^2}) \overset{.}{r}^2_{H}-4k (\frac{{t_p}^2}{{l_p}^2}) \overset{.}{r}^2_{H}}+(\frac{1}{1+z})^2 (\frac{{t_p}}{{l_p}}) \overset{.}{r_{H}}}}{2 (\frac{1}{1+z})^2 \overset{.}{r}_{H}}.
\end{array}
\end{equation}

The fine structure constant $\alpha$ does not change in time, for this reason, it is necessary to describe it as a parameter so that its variation with the expansion of the universe makes sense. The value of the change in the fine structure constant as a function of the expansion of the universe is taken as $\overset{\cdot }{\alpha }/\alpha =-1.7\cdot 10^{-18}$ (per second).

This expression for $\delta \alpha$ coincides well with that obtained in \cite{Lipovka_1}. In the Einstein-Cartan geometry framework, if we written it for the Riemannian manifold, when the cosmological constant $\Lambda = 0$, we have \cite{Lipovka_1}

\begin{equation}
\alpha=\frac{c^2}{32 \pi^2 G m} R.    
\end{equation}

Varying this expression we obtain \cite{Lipovka_2}

\begin{equation}
\delta \alpha=-\frac{H^3}{32 \pi^2 G m} \delta t.    
\end{equation}

This is a variation of the fine structure constant in time due to the change in curvature of the Riemannian manifold.

Now with this equation, it is possible to express a variation of the Hubble parameter $\delta H$ as a function of the variation of the fine structure constant $\delta \alpha$ presented in equation (92). To do this, it is necessary to use the variation of $\alpha$ that appears in the expressions (89) and (90). To be able to observe how the variation of the fine structure constant occurs as a function of the Hubble parameter:

a) $a r_{H} = 0$ and $a \overset{.}{r}_{H} T \neq 0$, we have

\begin{equation}
\begin{array}{cc}
\delta H= \frac{1}{t_p} \sqrt{(\frac{432 c^4 {m_{e}}^6 \epsilon_{0} \hbar G N}{96 \pi {m_p}^5 l_p A_{H} \rho_0 \alpha^6 \hbar^6})}\frac{(\frac{1}{1+z})^2 (\frac{\hbar {m_p} \delta \alpha}{{m_{e}} c}) - \sqrt{(\frac{1}{1+z})^2 (-8 \pi (\frac{1}{1+z})^2 (\frac{\hbar^2 {m_p}^2 \alpha}{{m_{e}}^2 c^2}) \delta \alpha (\frac{T}{T_p}) +(\frac{\hbar^2 
{m_p}^2 \delta \alpha^2}{{m_{e}}^2 c^2}) ((\frac{1}{1+z})^2-4k))}}{2 (\frac{1}{1+z})^2 (\frac{\hbar {m_p} \delta \alpha}{{m_{e}} c})}
\end{array}   
\end{equation}

and

b) $a r_{H} = 0$ and $a \overset{.}{r}_{H} T \neq 0$, the following holds

\begin{equation}
\begin{array}{cc}
\delta H= \frac{1}{t_p} \sqrt{(\frac{432 c^4 {m_{e}}^6 \epsilon_{0} \hbar G N}{96 \pi {m_p}^5 l_p A_{H} \rho_0 \alpha^6 \hbar^6})} \frac{\sqrt{(\frac{1}{1+z})^2 (-8 \pi (\frac{1}{1+z})^2 (\frac{\hbar^2 {m_p}^2 \alpha}{{m_{e}}^2 c^2}) \delta \alpha (\frac{T}{T_p}) +(\frac{\hbar^2 \delta \alpha^2}{{m_{e}}^2 c^2}) ((\frac{1}{1+z})^2-4k))}+(\frac{1}{1+z})^2 (\frac{\hbar^2 {m_p}^2 \alpha}{{m_{e}}^2 c^2})}{2 (\frac{1}{1+z})^2 (\frac{\hbar {m_p} \delta \alpha}{{m_{e}} c})}. 
\end{array}   
\end{equation}

It should also be considered that the apparent magnitude $\mu(z)$ is a measure of the brightness of a star or other astronomical objects. An object's apparent magnitude depends on its intrinsic luminosity, its distance, and any extinction of the object's light caused by interstellar dust along the line of sight to the observer.

There is an independent relationship between the luminosity distance $d_L (z)$ as a function of the redshift $z$ and the apparent magnitude $\mu$ is given by \cite{Barrientos}

\begin{equation}
\mu(z) = 5log [\frac{H_{0 d_L (z)}}{c}]-5logH(z)+B.    
\end{equation}

The light rays coming from the star usually go through an extinction phenomenon; caused by particles (dust) located in the path of the beam. Therefore, it is necessary to take into account the loss due to cosmic dust by introducing a correction factor $B$ in the magnitude, this factor moves the graph to the correct values, and in our calculations, such factor has the value of $B=59$, where $H_{0}$ is the Hubble constant evaluated at the present epoch $t_0$ and
$h := H_{0}/100$ $km/s/Mpc$ is the normalized Hubble constant, with the luminosity distance for a flat universe given by \cite{Barrientos}

\begin{equation}
\begin{array}{cc}
d_L(z)=\frac{c}{H_{0}}[z+\frac{1}{2}(1-q_{0})z^2-\frac{1}{6}(1-q_{0}-3q_{0}^2+j_{0})z^{3} & \\\\ + \frac{1}{24}(2-2q_{0}-15q_{0}^2-15q_{0}^3+5j_{0}+10q_{0}j_{0}+s_{0})z^4+...].
\end{array}
\end{equation}

Here $q_{0}$, $j_{0}$ and $s_{0}$ are the cosmographic parameters evaluated at the present epoch \cite{Barrientos}. The cosmographic parameters are obtained by performing a Taylor series expansion for the scale factor $a(t)$ as a function of time, thus

\begin{equation}
H(t)=\frac{1}{a} \frac{da}{dt},    
\end{equation}

\begin{equation}
q(t)=-\frac{1}{a} \frac{d^{2}a}{dt^{2}} \frac{1}{h^2},    
\end{equation}

\begin{equation}
j(t)=\frac{1}{a} \frac{d^{3}a}{dt^{3}} \frac{1}{h^3},     
\end{equation} 

\begin{equation}
s(t)=\frac{1}{a} \frac{d^{4}a}{dt^{4}} \frac{1}{h^4}   
\end{equation}
and 

\begin{equation}
l(t)=\frac{1}{a} \frac{d^{5}a}{dt^{5}} \frac{1}{h^5}.  
\end{equation}

However, the Dirac large numbers hypothesis is an observation made by Paul Dirac that relates the ratios of size scales in the Universe to force scales. The proportions constitute very large, dimensionless numbers: about $40$ orders of magnitude in the current cosmological epoch. According to Dirac's hypothesis, the apparent similarity of these relationships could not be a mere coincidence but could imply a cosmology with some unusual characteristics. These characteristic features are that the force of gravity, represented by the gravitational constant, is inversely proportional to the age of the universe ${\displaystyle G\propto 1/t\,}$;
the mass of the universe is proportional to the square of the age of the universe ${\displaystyle M\propto t^{2}}$; "physical constants" are not constants at present. Their values depend on the age of the Universe. Due to the above and the consideration that the universe is expanding, our interest arises in the variation of the fundamental constants with this cosmological expansion, this will be addressed in more detail in the next section.

\section{Comparison with Observational Data}

In this section, our interest is to compare the cosmological parameters such as the redshift $z$ and the temperature of the universe $T_U$, that our theoretical calculations show against astronomical data \cite{supernova_1} and \cite{supernova_2}.

To achieve this objective, the following cosmological parameters are considered: the temperature from the latest cosmic microwave background scattering display $T_{FCMW}=2.725 \pm 0.001$ $K$, the number of photons $N=1/(e^{(h\nu/KT)}-1) \sim 10^9$ photons $\sim T \sim (1+z)$, the time-dependent temperature of the universe $T(t)=T_0 (1+z)$, the scale factor of the FLRW metric $a(t)$ $\alpha$ $t^{1/2}$, the redshift is the variable to be plotted, age of the universe $t_U=(R_U/c)=10-15 \times 10^9$ years, the entropy per unit mass for the average density of matter in the universe $(S/M)=(1.1 \times 10^{16})/(\Omega h^2)$ $erg$ $K^-1$ $g^-1$ \cite{Vazquez}.

Using these data and equations (91) and (97), we wish to plot the value of the redshift $z$ against that of the Hubble parameter $H$.

The graph generated from equation (97), where $H$ $vs.$ $z$ is plotted in the Fig. $(1)$. For this, the following data reported in CODATA \cite{CODATA} are used $c=299792458$ $m/s$, $\pi=3.141592654$, $N=8 \times 10^{88}$ bits, $h=6.626070040$ $J$ $\cdot$ $s$, $G=6.67408$ $\times$ $10^{-11}$ $m^3$ $kg^-1$ $s^-2$, $m_{e}=9.10938356 \times 10^{-31}$ $kg$, $\epsilon_{0}=8.854187187 \times 10^{-12}$ $Fm^-1$, $R_{U}=4.4 \times 10^{26}$ $m$, $|\vec{k}|=0.008975979$ $rad/m$, $r_{H}=1.5672\times 10^{27}$ $m$, $T_{U}=2.725$ $K$, $\rho_{c}=\rho_{0}=1.8788*10^{-26}$ $h^2 kg m^-3$, $H_{0}=68.37$ $km/s/Mpc$, $z=20$, $\alpha=7.2973525693*10^{-3}$, $A_{H}=7.716115 \times 10^{54}$ $m^{2}$, $k=1$, $q_{0}=-0.55$ and $j_{0}=1$.

\begin{center}
\includegraphics[scale=.55]{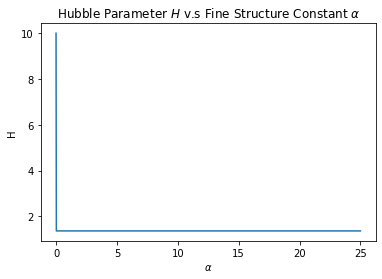}
Fig. $1$. Apparent magnitude $\mu(z)$ vs. redshift $z$ diagram in comparison with \cite{Barrientos}.
\end{center}

In this graph, it is possible to see how the apparent magnitude $\mu(z)$ grows with the redshift $z$.

Likewise, considering equation (95) it is possible to observe the change in the fine structure constant $\alpha$ as a function of the change in the Hubble parameter $H$, this can be seen in Fig. (2).

\begin{center}

\includegraphics[scale=.55]{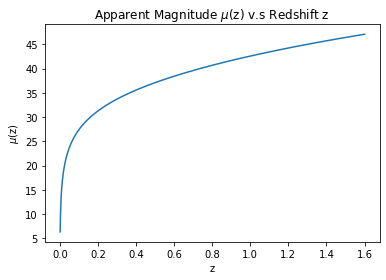}

Fig. $2$. Hubble parameter $H$ vs. fine structure constant $\alpha$, cf. \cite{Kant}.
\end{center}

In the following sections, a more detailed discussion of this result and the conclusions will be given.

\section{Discussion}

In this work, the FLRW metric was used for the Klein-Gordon equation. This represents a scalar field from which the creation and destruction (entering and leaving the cosmological horizon) of particles are observed, and which is located on the surface of the observable universe. By using the Klein-Gordon equation it is possible to find a wave equation, where the normalization of the solution of this equation allows us to represent the temperature of the universe. This temperature of the universe is dependent on the electromagnetic field present throughout the universe. For this reason, it was considered the 4-electromagnetic momentum for such a field. Due to the expansion of the universe, the electromagnetic field in it also expands. Therefore with this temperature of the universe, it was possible to find the Hubble parameter. In that sense, we could make a comparison of this Hubble parameter $H(z)$ vs. the apparent magnitude $\mu(z)$ and plot it.

Now, in Fig. 2 we can see a curve that describes the variation of the fine structure constant $\alpha$ as a function of the Hubble parameter $H$. This is compared with Fig. 2 from the work of Kant and Chand \cite{Kant}. This figure plots the Hubble parameter $H$ as a function of time $t$.
To analyze what was mentioned in the previous paragraph, equation (52) described in \cite{Lipovka_3} will be used

\begin{equation}
(\textit{R}-4 \Lambda) \frac{c^{4}}{16 \pi G} = 2 \pi m_{e} c^{2} \alpha.     
\end{equation}

Also in equation (30) described in \cite{Lipovka_1}

\begin{equation}
\delta \alpha = \frac{(1- \alpha^2)^{3/2} H c^3}{8 \pi^2 mcG} (\textit{R}-4 \Lambda) \delta t.    
\end{equation}

In equation (104) the scalar curvature \textit{R} can be written in terms of the Hubble parameter $H$. It can be seen that the fine structure constant $\alpha$ has a behavior directly proportional to the Hubble parameter $H$.
In equation (105) it can be seen that the fine structure constant $\alpha$ is directly proportional to time $t$. For this reason, it is possible to compare the Fig. 2 in the present paper with Fig. 2 in \cite{Kant}.

In Fig. 1 of \cite{Barrientos}, the apparent magnitude $\mu$ is shown as a function of the redshift $z$. This result is consistent with our Fig. 1 of that apparent magnitude. In Barrientos et al \cite{Barrientos} an expanding universe was considered in which dark matter is not necessary and is consistent with astronomical observations. This could even give us a way forward to enhance studies within the framework of MOND.

Furthermore, the number of bits $N$ plays a fundamental role in our equation, suggesting that the holographic principle is very important in the emergence of the scalar field on the surface of the universe.

We also consider that it is possible to apply the Klein-Gordon equation in curved space-time to the Schwarzschild and Reissner-Nordstrom metrics. In this way, it would be possible to find the temperature for Hawking radiation and the laws of black hole thermodynamics. This would represent a different path to obtain these laws in a different way them Hawking's. This would underline the importance of the contribution of the scalar field given by the Klein-Gordon equation to the temperature of both the universe and the black holes.

We conclude that the expansion of the universe contributes to the variation of the fundamental constants and in turn to the variation of the temperature and entropy of the universe. This is reflected in the fact that there may be a relationship between the number of bits $N$ and the variation of the fundamental constants, two important objects of study in quantum gravity theories that attempt to unify all fundamental forces.

The wave function for the universe is given by the Wheeler-DeWitt wave equation. In the Klein-Gordon equation in curved space-time given in (7) and developed in equations (8) and (9); we have used the approximation for the wave function corresponding to the wave equation of the scalar field. However, although this equation is written in curved spacetime for the FLRW metric, the wave function does not contain all the information for the universe. For this reason, we consider that it is more appropriate to use the wave function given by the Wheeler-DeWitt equation in the future. The use of the Wheeler-DeWitt equation is an alternative formulation in comparison with the quantization of the Klein-Gordon field, because with the wave function of the Wheeler-DeWitt the quantization is for the universe, and with the Klein-Gordon equation it is the quantization of a field. 

We also consider that it is interesting to couple the Wheeler-DeWitt equation and the Klein-Gordon equation in the FLRW metric, to find a value of the wave function of the universe, which is a function of the scalar field in the cosmological horizon. The quantum effects of the electromagnetic field are described for the Wheeler-DeWitt equation $H | \psi \rangle=0$.

Due to what was commented on in the previous paragraph, when considering the wave function explicitly, the semiclassical approximations for large Dirac numbers and a relativistic classical electromagnetic field and momentum can no longer be considered. In that case, we must start from the Ward-Takashi equation given in equation (56). Furthermore, in a cosmological context, the Ward-Takashi equation must be written in a curved space-time and its corresponding wave function must contain all the information of the universe. For this reason, it is more appropriate to use the wave function described by Wheeler and Dewitt.

In various works on quantum gravity \cite{Duff_1}, the entropy for black holes has been calculated in terms of qubits. Likewise, since the Klein-Gordon equation has been used, we consider that the use of “quantum bits” or “qubits” is more appropriate for describing the entropy of the universe in terms of information theory.

\section{Conclusions}

In this work, it was observed that it is possible to consider that a scalar field given by the Klein-Gordon equation in curved space-time and the electromagnetic field contribute to the temperature of the universe for the cosmological horizon.

The presence of these scalar and electromagnetic fields in the expansion of the universe has important consequences. This is done by incorporating the hypothesis of large Dirac numbers, for this reason it was possible to find an expression for the Hubble parameter as a function of the temperature of the universe, the number of bits $N$ for the universe, and the fundamental constants.

It could be observed that the expansion of the universe contributes to the change of the fundamental constants, as well as its number of bits $N$.

As future work, we consider it interesting to further study the implications and consequences of the number of bits in cosmology and the expansion of the universe, in addition to what deeper aspects of the holographic principle can be addressed.

On the other hand, in his article Duff \cite{Duff_1} describes how the entropy of a black hole can be expressed in terms of qubits instead of bits, we consider that it is of interest to understand how the information expressed in qubits for the entire universe can be considered a fundamental entity.

Furthermore, there is a formulation made by Frieden \cite{Frieden} where he derives the laws of physics through information theory, thanks to constructing a Lagrangian in informational terms, and through the Euler-Lagrange equations, he obtains the most important equations of the laws of physics. It is natural to address the question of whether there is a relationship between our work, and Frieden's approach, because a derivation of the laws of physics from the laws of information theory seems to be fundamental.

We intend to derive from informational principle a Lagrangian that considers the aforementioned coupling of the Klein-Gordon equation in the FLRW metric and the Wheeler-DeWitt equation.

\bigskip

$\,$

$\,$

\break

\end{document}